\documentclass[12pt]{article}

\usepackage{graphics,psfrag}
\usepackage{amsthm,amssymb,epsfig,amsmath,euscript,array,cite}

\newcommand{\be}{\begin{equation}}
\newcommand{\ee}{\end{equation}}
\def\bea{\begin{eqnarray}}
\def\eea{\end{eqnarray}}
\newcommand{\eq}[1]{(\ref{#1})}

\def\gh{{\hat\gamma}}

\newcommand{\beq}{\begin{equation}}
\newcommand{\eeq}{\end{equation}}
\newcommand{\ben}{\begin{eqnarray}}
\newcommand{\een}{\end{eqnarray}}
\newcommand{\bes}{\begin{subequations}}
\newcommand{\ees}{\end{subequations}}
\newcommand{\blg}{\begin{align}}
\newcommand{\elg}{\end{align}}

\newcommand{\cN}{{\cal N}}

\newcommand{\hg}{{\hat{\gamma}}}

\newcommand{\startappendix}{
\setcounter{section}{0}
\renewcommand{\thesection}{\Alph{section}}}

\newcommand{\Appendix}[1]{
\refstepcounter{section}
\begin{flushleft}
{\large\bf Appendix \thesection: #1}
\end{flushleft}}

\def\one{\mbox{1 \kern-.59em {\rm l}}}

\def\a{\alpha}      
\def\b{\beta}       
\def\g{\gamma}    
\def\d{\delta}  \def\D{\Delta}  
\def\e{\epsilon}

\def\k{\kappa}
\def\l{\lambda} 
 
\def\o{\omega}
 
\def\r{\rho}
\def\s{\sigma}  
\def\t{\tau}
\def\th{\theta}

 \def\cE{{\cal E}} 
 \def\cH{{\cal H}} 
\def\cJ{{\cal J}}  
 \def\cN{{\cal N}}

 \def\cZ{{\cal Z}}

\begin{document}

\hfill{WITS-CTP-057}

\vspace{20pt}

\begin{center}

{\Large \bf
Semiclassical strings in marginally deformed toric AdS/CFT
}
\vspace{20pt}

{\bf
 Dimitrios Giataganas
}

{\em
National Institute for Theoretical Physics,\\
School of Physics and Centre for Theoretical Physics,\\
University of the Witwatersrand,\\
Wits, 2050,\\
South Africa
}

{\small \sffamily
dimitrios.giataganas@wits.ac.za
}

\vspace{30pt}
{\bf Abstract}
\end{center}

We study string solutions in the $\beta$-deformed Sasaki-Einstein gauge/gravity dualities. We find that the BPS point-like strings move in the submanifolds where the two $U(1)$ circles shrink to zero size. In the corresponding $\mathbb{T}^3$ fibration description, the strings live on the edges of the polyhedron, where the $\mathbb{T}^3$ fibration degenerates to $\mathbb{T}^1$. Moreover, we find that for each deformed Sasaki-Einstein manifold the BPS string solutions exist only for particular values of the deformation parameter. Our results imply that in the dual field theory the corresponding BPS operators exist only for these particular values of the deformation parameter we find. We  also examine the non-BPS strings, derive their dispersion relations and compare them with the undeformed ones. Finally, we comment on the range of the validity of our solutions and their dependence on the deformation parameter.

\setcounter{page}0
\newpage

\section{Introduction}

During the last years there is a lot of work done in the original AdS/CFT \cite{tHooft74,maldacena98,witten98} correspondence in maximally supersymmetric theories. More recently similar ideas have been applied
to theories with less supersymmetries. These reduced supersymmetric gauge/gravity dualities can be obtained either by deforming theories with more supersymmetries, or by constructing them directly with appropriate handling of the Dp-branes. Recently there is a lot of progress in both of these directions, and several gauge/gravity dualities with less supersymmetries have been found.

More precisely, by placing multiple D3-branes to the singularity of the metric cones over the Sasaki-Einstein manifolds we obtain a gauge/gravity duality that preserves $\cN=1$ supersymmetries
\cite{gauntlett04a,gauntlett04b,gauntlett06,martelli04}. The gravity background is $AdS_5\times Y^{p,q}$ and the dual quiver field theory can be constructed using the fact that the Calabi-Yau cone of these manifolds is toric. The gauge theories \cite{benvenuti04} have a product gauge group $U(N)\times U(N)$, and hence bifundamental matter superfields, where two of them transform in the $(N,\bar{N})$ and the other two in the $(\bar{N},N)$ representations.

On the other hand, there is an extended use of the $\beta$ deformations \cite{leigh95}. There is a known way to marginally deform  both a field theory and the corresponding dual background with at least $U(1)\times U(1)$ global symmetry. In the resulting theory the conserved supersymmetry is less or equal to the initial one, or even there is not at all \cite{LM05,frolov05}.

In this paper we are interested in the $\beta$-deformed Sasaki-Einstein \cite{LM05,benvenutibeta,lee} gauge/ gravity dualities. These backgrounds have three $U(1)$ isometries and hence the $\beta$ deformation can be done, either by keeping untouched the amount of supersymmetry preserved or by reducing it to zero. How much supersymmetry preserved, depends on which angles the TsT transformation is done. Moreover, if the deformation is multiple, where three TsT transformations are done, the supersymmetry is broken completely. More details on the $\beta$ deformations in these backgrounds are given in the section 2.

In the Sasaki-Einstein gauge/gravity dualities, there are several papers examining string and brane solutions. For the special case of  $AdS_5\times T^{1,1}$, the semiclassical string solutions and their dual field theory operators were examined in \cite{kim1,pons1,schvellinger1}, and the pulsating string solutions in \cite{rashkov06}. Giant magnons
and spiky strings moving in a sector of $AdS_5\times T^{1,1}$ have been examined in \cite{benvenuti08}. In the whole class of Sasaki-Einstein manifolds the BPS string solutions \cite{benvenutiY05,giataganas3} and their dual long BPS operators have been also examined \cite{benvenutiY05}. Similar analysis applied to the generalized spaces with cohomogeneity two, called $L^{p,q,r}$ \cite{cveticL05}, in \cite{benvenutiL05}. More recently, it has been done an extended analysis of the non-BPS string solutions in these backgrounds \cite{giataganas3,giataganas4}. A discussion on the pulsating strings is presented in \cite{rashkov07}. Moreover, the dual giant gravitons have been studied in \cite{martelli2}.

In the $\beta$ deformed backgrounds the giant gravitons and the dual giant gravitons have been extensively examined in \cite{zaffaroni07}. There has been found that the BPS dual giant gravitons, live on the edges of the corresponding rational polyhedron where the $\mathbb{T}^3$ fibration degenerates to $\mathbb{T}^1$. Furthermore, in \cite{rashkovT09} the giant magnons and single spike solutions have been examined in the special case of the deformed conifold.

A natural extension of these studies, is to examine the properties of the string solutions in these $\beta$ deformed spaces\footnote{In the deformed sphere the problem is already discussed extensively in \cite{tseytlinbeta}. For some collective work on strings solutions in AdS/CFT one can look at \cite{tseytlinrev} and references inside.}.
The first problem we solve, is to find the BPS string solutions. It turns out that the BPS strings are allowed to move in a very constrained space. They live in the submanifold where two of the $U(1)$ circles shrink to zero size and the corresponding two torus collapses. In the $\mathbb{T}^3$ fibration description this corresponds to the edges of the canonical polyhedron. Moreover, it appears that there is a  relation between the parameter $a(p,q)$ which specifies the exact manifold $Y^{p,q}$, and the deformation parameter. That means that a BPS string solution in a specific $Y^{p,q}$ where the $p$ and $q$ are fixed, exist only for a particular value of the deformation parameter $\gh$. In other words for an appropriately chosen value of the deformation parameter $\gh$, in a way that all the Sasaki-Einstein constrains are satisfied, the specific $Y^{p,q}$ where the BPS string live has been already chosen. This statement is a result of the equation \eq{bg}.

Then we examine the properties of some non-BPS string solutions, and compare them with the undeformed ones. 
There exist solutions that are valid only for $\gh \neq 0$ and although at the limit $\gh \rightarrow 0$ the undeformed result is reproduced, there are other cases that the $\gh$ is bounded from below by a finite value. This bounding is enforced in order to get a real solution. Actually even in these cases at the limit $\gh \rightarrow 0$ the complex undeformed solution is reproduced, which however is not acceptable in the undeformed theory. Hence in the deformed space there exist solutions that are not acceptable in the undeformed one, and this is because the deformation parameter provides more freedom to satisfy the equations of motion, the Virasoro constrains and the Sasaki-Einstein constraints. We also calculate the dispersion relations of these configurations and comment on their properties.

The paper is organized as follows. In the section 2 we present the deformed Sasaki-Einstein background and briefly mention the convenient methods available that can be used to $\beta$-deform these backgrounds. This section is supported by the Appendix A, where we derive the equations of motion, the Virasoro constraints and the corresponding conjugate momenta. In section 3 we study the BPS string solutions in the deformed Sasaki-Einstein backgrounds. In section 4, the non BPS string solutions are analyzed. In the final section we present the concluding remarks and give further directions. In Appendix B we briefly discuss a Hamiltonian approach to look at some properties of the BPS string solutions in marginally deformed toric geometries. In Appendix C, we discuss the $\gh\rightarrow 0$ limit for the solutions of general a $\beta$-deformed background. In Appendix D, there are some useful formulas of the Sasaki-Einstein backgrounds that we use in the main paper.

\section{Deformed $Y^{p,q}$ Backgrounds and their dual gauge theories}

There are several equivalent methods that can be used to  derive  the $\beta$ deformed $Y^{p,q}$ backgrounds from the original ones. These technics can be applied to even more general backgrounds, that have at least a $U(1)\times U(1)$ global symmetry in the internal space, and does not even matter if these spaces are conformal or non-conformal since the  deformations we are interested here are applied only in the internal space. In the field theory side, the corresponding global symmetry is used to construct the deformed dual field theory. Using the $U(1)$ charges of the fields, the ordinary multiplication between the fields is redefined to a version of star product in order to get the dual deformed field theory. This redefinition just introduces some phases in the lagrangian.

In the gravity side one can construct the solutions by reducing the ten dimensional theory to eight dimensions on the two torus. Then the eight dimensional theory, at the level of supergravity, is invariant under the group $SO(2,2,R)\simeq SL(2,R)_\t\times SL(2,R)_\r$. This symmetry act on the K\"{a}hler modulus $\t$ and the complex structure modulus $\r$ of the internal torus respectively.

The important role in the construction of the deformed backgrounds plays the $SL(2,R)_\t$ symmetry acting on the K\"{a}hler modulus which has as a real part  the B-field and imaginary the volume of the two torus. By acting this symmetry on $\t$ and choosing appropriately the $SL(2,R)$ parameters, where three of them are finite numbers and the remaining one is the deformation parameter $\gh$, one can generate the Lunin-Maldacena backgrounds.

Obviously, when the transformation parameter is an integer the new background coincides with the old one. This can be also seen by noticing that integer $\gh$ corresponds to using the discrete $O(2,2,\mathbb{Z})$ group acting on the moduli space of a two dimensional world-sheet conformal field theory, which results a conformal field theory physically equivalent to the  initial one.

Maybe the most practical way to $\beta$-deform a gravity background is the TsT deformation. This consists of a T-duality along an appropriately chosen $U(1)$ direction followed by a shift on an angle that corresponds to another $U(1)$ direction and where the deformation parameter enters, and finally a T-duality on the initial angle.

In $S^5$ dual background the choice of the angles obviously does not play role and the resulting deformation leads to $\cN=1$ supersymmetric dual background. This is not the case in the Sasaki-Einstein gauge/gravity dualities. In order to marginally deform the $\cN=1$ supersymmetric  background, without reducing further the supersymmetry one has to choose the angle which when shifted leave invariant the holomorphic $(3,0)$ form $\Omega$, which specify the Calabi-Yau structure on the metric cone\footnote{Together with the K\"{a}hler form.} \cite{martelli04}. This means that the Killing spinors remain also invariant.

A transformation along the angle $\psi$, which is the azimuthal coordinate on the axially squashed $S^2$ sphere as can be seen in metric \eq{sasakiundef}, modifies the holomorphic 3-form. Hence we perform the T-dualities and the intermediate shift, along the other two $U(1)$ directions, which correspond to the angles $\a$ and $\phi$ in the metric \eq{sasakiundef}, and leave the number of preserved supersymmetry unchanged. The supergravity description is valid in the limit of small curvature and when $R\b\ll 1$, with $\gh:= R^2\b$. Where  $\gh$ and  $\b$ are the deformation parameters appear in supergravity background and in the dual field theory.

The deformed background \eq{sasakiundef} can be written as \cite{LM05,ozer}
\ben
\label{gse1}\nonumber
  d s^2 &=& \frac{1-y}{6}d\theta^2+\frac{1}{w(y) q(y)}dy^2+G\bigg[\left(\frac{1-y}{6}\sin^2\theta+\frac{q(y)}{9}\cos^2\theta +w f^2 \cos^2\theta\right)
      d\phi^2\\\nonumber &+& w(y) d\a^2 + \Big(\frac{q(y)}{9} +w f^2
       + \gh^2 h(y,\th)\Big)d \psi^2-2\left(\frac{q(y)}{9} +w f^2\right)\cos\th d\phi d\psi\\&&\qquad\qquad\qquad\qquad\qquad\qquad\qquad
       \qquad~+ 2 w f d\a d\psi  -2 w f \cos\th d\a d\phi\bigg]
\een
where
\ben
w(y)  =  \frac{2(a-y^2)}{1-y}~ ,\quad
q(y)  =  \frac{a-3y^2+2y^3}{a-y^2}~ , \quad
f(y)  =  \frac{a-2y+y^2}{6(a-y^2)}~.
\een
The metric is written in a form, where the deformation effects can be easily seen\footnote{In our case, the
metric is just modified as $g_{ij}\rightarrow G g_{ij}$ and $g_{33}\rightarrow G (g_{33} +\hg^2 h(y,\th))$. Where $g_{ij}$ are the metric elements in the three $U(1)$ directions.}.
Under the TsT transformation a non trivial B-field is generated which is given by
\ben
B = G R^2(B_{\a \phi} d\a \wedge
d\phi + B_{\a \psi} d\a \wedge d\psi + B_{\phi \psi} d\phi \wedge d\psi)
\een
with
\be\label{bcomponents}
B_{\a\phi} = \hat{\g} Z~, \quad
B_{\a\psi} = -\hat{\g}\frac{w q}{9} \cos\th~,\quad
B_{\phi \psi} =  -\hat{\g}\frac{f (a-y^2) }{3}\sin^2\th\, .
\ee
Where the function $G(y,\th)$ is equal to
\ben
G(y,\th)=\frac{1}{1+\hg^2 Z(y,\th)}
\een
with
\ben
Z=\frac{2 q(y) \cos^2\th+ 3 (1-y) \sin^2\th}{18} w(y)\quad\mbox{and}\quad h(y,\th)=\frac{\sin^2\th}{27}q(y)(a-y^2)~.
\een
From now on for convenience we avoid to write the arguments of the functions.

The parameter $a$ in the above equations is related to $p$ and $q$ from \eq{apq}, and is restricted to the range
\be\label{acon}
0<a<1~.
\ee
In the undeformed background \eq{sasakiundef} to make  the base
$B_4$ of the space, an axially squashed $S^2$ bundle
over the round $S^2$ we choose the ranges of the
coordinates $(\theta,\phi,y,\psi)$ to be $0\le\theta\le \pi$,
$0\le \phi \le 2\pi$, $y_1\le y\le y_2$ and $0\le \psi \le 2\pi$. Before the deformation  applied, the parameter $\psi$ is the azimuthal coordinate on the axially
squashed $S^2$ fibre and the round sphere $S^2$ is parameterized by $(\theta,\phi)$.
To determine the poles of the axially squashed $S^2$ sphere we need to solve the equation $q(y)=0$, which is cubic and has three real roots, one negative and two positive. Naming the negative root $y_{q-}$ and the smallest positive root $y_{q+}$ we must choose the range of the coordinate $y$ to be
\be\label{yqy}
y_{q-}\leq y\leq y_{q+}~,
\ee
with the boundaries corresponding to the south and north poles of the squashed $S^2$ fibre.

It should also be mentioned that when one examines BPS objects in these spaces usually does not need all the information for the background. Actually one can even consider the general case of a toric  Calabi-Yau cone and study supersymmetric quantities without
to know the explicit form of the metric. It can be shown \cite{martelli04} that the metric of the Calabi-Yau cone can be written as
\be\label{metriccy}
ds^2=g_{ij} dx^i dx^j +g_{ij} d\phi^i d\phi^j~,
\ee
where the manifold is described by the complex coordinates $z^i=x^i+i \phi^i$ and the K\"{a}hler metric is related to the  K\"{a}hler
potential $F(x^i)$ by
\be
g_{ij}=\frac{\partial^2 F}{\partial x^i \partial x^j}~.
\ee
By doing the TsT transformation to the angles $\phi_1$ and $\phi_2$ of \eq{metriccy} and leave the holomorphic three form invariant,
we get the new deformed background where the metric deforms in the way presented
in the footnote 3.

The  quiver gauge theories are deformed in a similar way with the $\cN=4$ super Yang-Mills. The ordinary multiplication between the fields is replaced by a star product introducing a phase in the multiplication:
\ben
f\star g\equiv f g \,e^{i \pi \b( Q_1^f Q_2^g-Q_2^f Q_1^g)}~,
\een
where $Q$ are the charges of the matter fields under the two $U(1)$ global symmetries. This results to a deformed superpotential of the following form
\be
A_1 B_1 A_2 B_2 e^{i \pi \b}-A_1 B_2 A_2 B_1  e^{-i \pi \b}~.
\ee

\section{BPS Solutions}

In this section we find the massless geodesics that satisfy the BPS condition in the deformed Sasaki-Einstein backgrounds\footnote{For this section as well as for the next ones we use the equations and relations derived in Appendix A.}.
In order to do that we consider a massless point-like string localized at a point
in the bulk of  $AdS$ with global time $t=\k \t$, which is moving along all the
directions in the internal manifold. The relevant ansatz describing this motion is
\be
\a=\a(\t),\qquad\phi=\phi(\t),\qquad\psi=\psi(\t),\qquad\theta=\theta(\t)\qquad\mbox{and}\qquad y=y(\t)~.
\ee
We see from the action \eq{actiond} that there is no B-field contribution since only  time derivatives enter
in the equations.
In this case the second Virasoro constraint \eq{vc2} can be written in the following form
\be\label{k11}
\l \k^2=\frac{6}{1-y}J_\th^2 + w q J_y^2+\frac{1}{G w}J_\a^2+ \l \k_1
\ee
where
\ben
J_y =\sqrt{\l} \frac{1}{w q} \dot{y},\quad
J_\theta =\sqrt{\l} \frac{1-y}{6} \dot{\theta},\quad
J_\a =\sqrt{\l}G(w \dot{a}-  w f c_\theta \dot{\phi} +w f \dot{\psi})
\een
and $\k_1$ is defined as
\be
\k_1:= G\left(\left(\frac{1-y}{6} s_\th^2+\frac{q}{9} c_\th^2\right) \dot{\phi}^2 +
\left(\frac{q}{9} + \gh^2 h\right)\dot{\psi}^2-2 \dot{\phi }\dot{\psi} \frac{q}{9} c_\th \right)~.
\ee
We would like to write this expression in terms of the momenta and
the R-charge in a way that is positive definite so to find the BPS bound and hence the BPS solutions that saturate it.
After some algebra, $\k_1$ can be written as
\ben\nonumber
\k_1&=&\frac{1}{\l G}\left(\frac{6}{(1-y) s_\th^2}\left(J_\phi+c_\th J_\psi\right)^2  + \frac{9}{ q}\left(J_\psi- f J_a\right)^2\right) \\
&&\qquad \qquad \qquad \qquad - G \gh^2 h \dot{\psi}^2\left(1+\gh^2 h\left(\frac{9}{q}+\frac{6}{1-y}\cot^2\th\right)\right)~.
\een
By substituting $G$ in the last term, the expression is simplified to
\be
\k_1=\frac{1}{\l G}\left(\frac{6}{(1-y) s_\th^2}(J_\phi+c_\th J_\psi)^2  + \frac{9}{ q}(J_\psi- f J_a)^2 \right)
-\gh^2 h \dot{\psi}^2~.
\ee
So the equation \eq{k11} which is directly related to the energy of the string and hence the conformal dimension of the dual operator can be written in terms of the  R-charge\footnote{$Q_R=2 J_\psi-\frac{1}{3} J_\a$.} as
\ben\label{kk2}\nonumber
\l\k^2&=& w q J_y^2+\frac{6}{1-y}\left(J_\th^2+\frac{1}{G s_\th^2}(J_\phi+c_\th J_\psi)^2\right)+\frac{1}{G}\left[(\frac{3}{2} Q_R)^2+\frac{1}{w q}( J_\a+3 y Q_R)^2\right]\\ && \hspace{10.8cm} -\l\gh^2 h \dot{\psi}^2~.
\een
This expression reproduce the undeformed result in the limit $\gh=0$ as expected. Inside the region where the parameters and the coordinates defined, the function $h$ is positive. Hence the last term has to be combined with other terms in order the RHS of the above equation to become positive definite. To do that one has to substitute the expression for $G$ and examine the $\gh^2$ order terms separately.
The first step is to express $\dot{\psi}$ appropriately in terms of the momenta
\be
\sqrt{\l}\dot{\psi}=\frac{9}{q}\left(\left(J_\phi+c_\th J_\psi\right)\frac{2 q c_\th}{3(1-y) s_\th^2}+(J_\psi-f J_\a)\right)~.
\ee
Substituting $\dot{\psi}$ in \eq{kk2} and after some algebra we manage to group together the terms in a nice way
\ben\nonumber
\l\k^2&=&(\frac{3}{2} Q_R)^2 +\frac{1}{w q}(J_\a +3 y Q_R)^2+ \frac{6}{1-y}\left(J_\th^2+\frac{1}{s_\th^2}\left(J_\phi+c_\th J_\psi\right)^2\right)
+ w q J_y^2 \\\label{kfinal}
&&\qquad\qquad\qquad\qquad\qquad\qquad\qquad\qquad+ \gh^2\left(\frac{Z}{w}J_\a^2+w \left(J_\phi+ f c_\th J_\a\right)^2\right)~.
\een
The functions $w(y),\,q(y)$ and $Z(y,\th)$ are positive in the range we are interested, so each term in the previous expression is positive.
The conformal dimension $\D$ of the dual operator to the string configurations we consider, is equal to string energy $E$ given by \eq{1e}. 
In order to get the BPS solutions from \eq{kfinal} we have to saturate the bound $\D=3/2 \,Q_R$, and this happens only when
\be\label{bpseq1}
J_y=0~, \qquad J_\th=0~,\qquad J_\phi=-c_\th J_\psi~,\qquad J_\a=-3 y Q_R~,
\ee
which are the same conditions as in the undeformed case and additionally the following equations must hold
\be\label{bpseq2}
Z=0~\quad\mbox{or}\quad J_\a=0~,\qquad\mbox{and}\qquad J_\phi=- f c_\th J_\a~.
\ee
Of course we consider $\gh\neq 0$, otherwise we will end up to the already examined undeformed case \cite{benvenutiY05,giataganas3}.

Let us start by supposing that $J_\a=0$. Then the last equation of \eq{bpseq1} gives
\be
y=0~.
\ee
This value of $y$ leads to important simplifications and although this is a limiting example, since it is a very special point in our manifold we examine if the full solution exist for completeness. The third equation of \eq{bpseq1} and  the last equation of \eq{bpseq2} using the fact that $J_\a=0$ gives
\be
\dot{\phi}=0~,\qquad \th=\frac{\pi}{2}~,\qquad\dot{a}=-\frac{\dot{\psi}}{6}~.
\ee
It can be easily seen that this solution satisfies the equations of motion, providing that $\ddot{a}=0$, and hence it is acceptable. However as we already noticed it is not useful due to the fact that is restricted only to $y=0$.

More interesting things happen when we minimize \eq{kfinal} by requiring
\be\label{D}
Z=0~.
\ee
This is a very strong condition by itself, and constrain significantly $\th$ as well as $y$. By solving \eq{D} for $y$ in terms of $a$ and $\th$ we get three solutions which are generally outside the area bounded by $y_{q-}(a),\,y_{q+}(a)$ except when $\th$ or $a$ take some very special values.
By doing the analysis we find that the only acceptable solutions are these for $\th=0,\,\pi$ where $y$ becomes equal to $y_{q\pm}$ for any $a$.\footnote{There exist more special solutions for $a=1$ and $y=1$, for any value of $\theta$ but these are not interesting.}
In Figure 1 we can also see that the acceptable solutions are:
\begin{figure}
\centerline{\includegraphics[width=65mm]{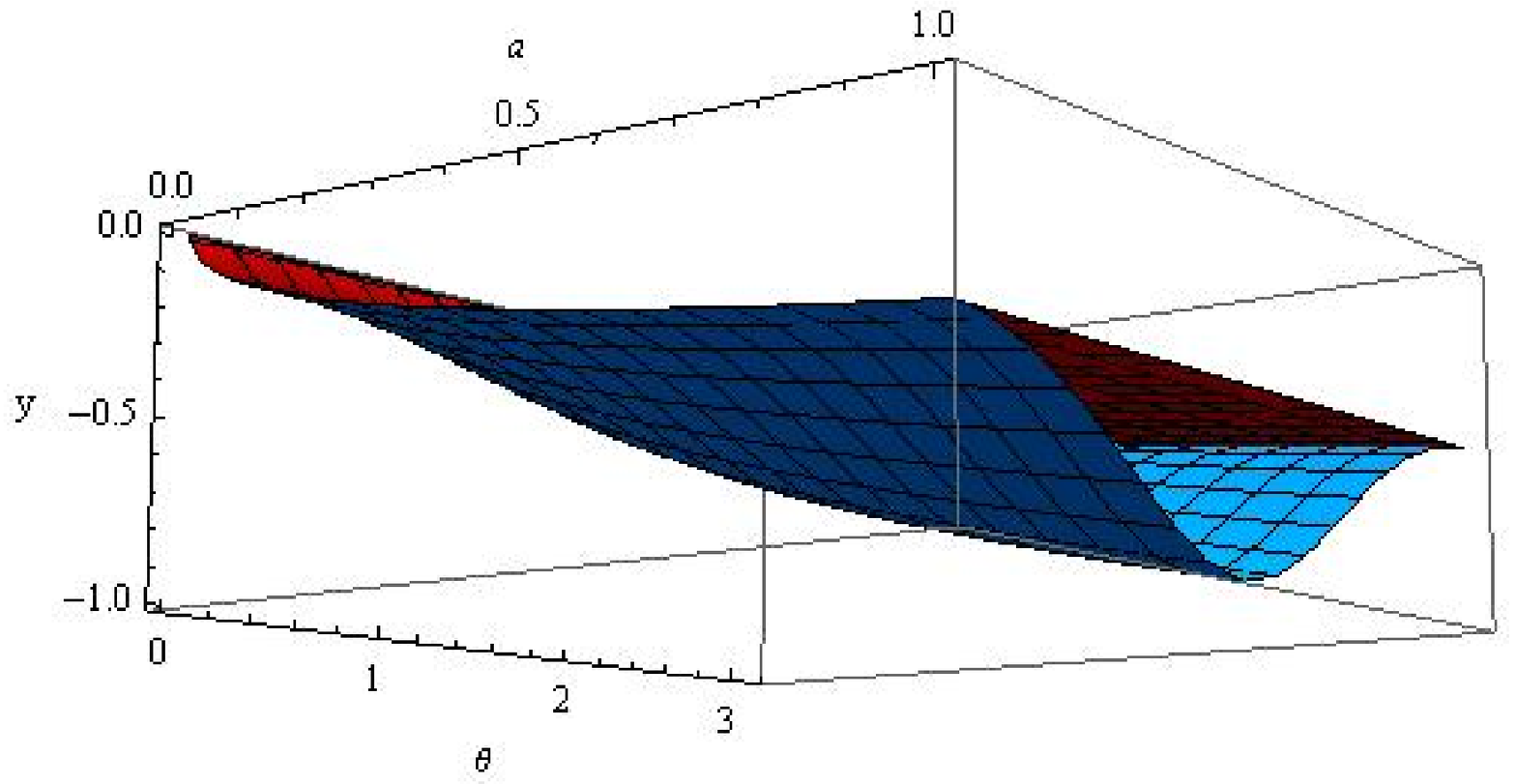} \quad\quad \includegraphics[width=65mm]{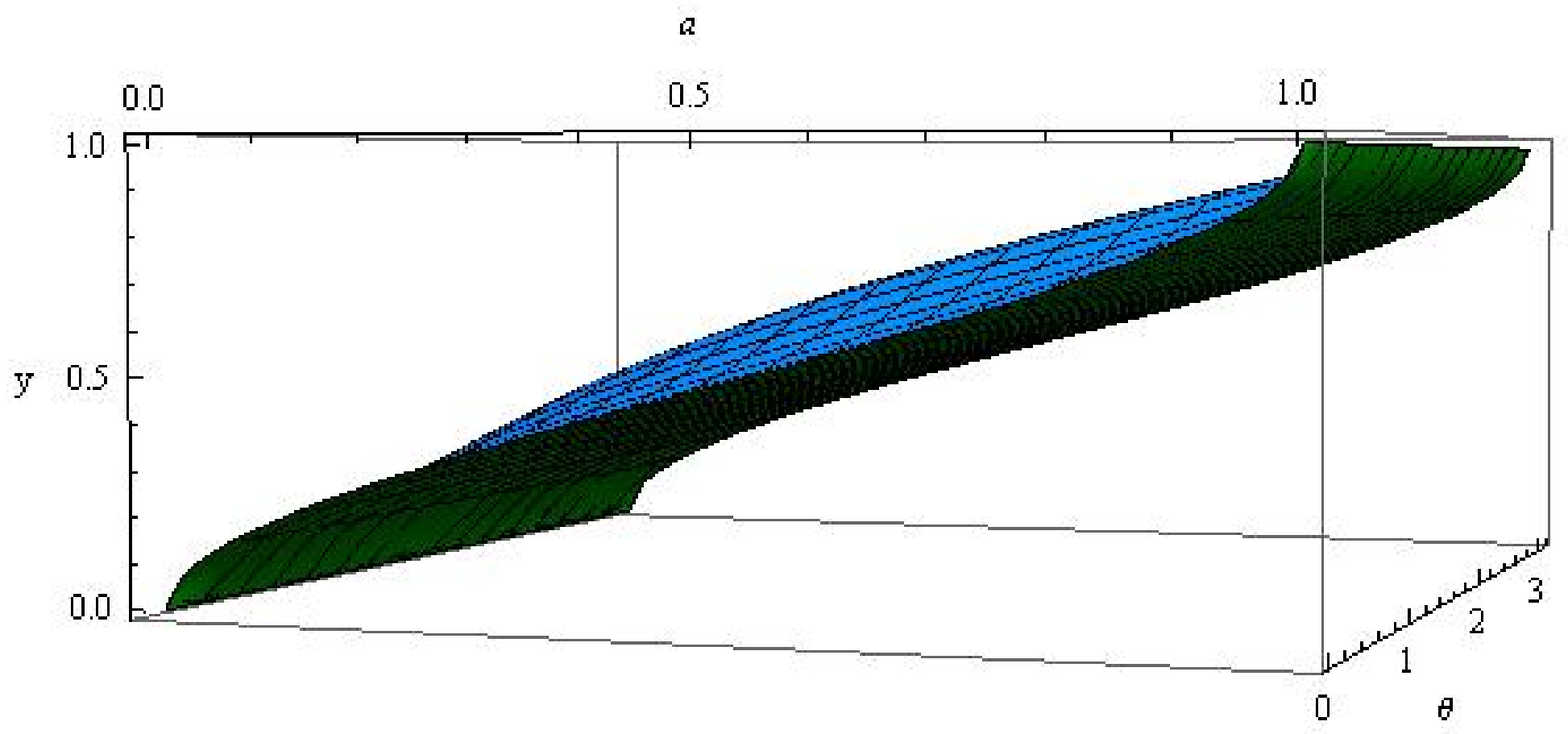}}
\caption{\small{Two of the solutions of $Z=0$ are plotted. In the first plot we see the $y_{q-}$ (red) and the solution $y$ (blue) in terms of $a$ and $\theta$. We can only saturate the equality of the condition \eq{yqy} to $y_{q-}$ for $\th=0,\,\pi$. In the second plot, we draw $y_{q+}$ (green) and the other solution of \eq{D}. The solution $y$ is greater than $y_{q+}$ and we can only saturate the equality for $\th=0,\,\pi$.}}
\end{figure}
\be\label{equationsbps}
\theta=0,\, \pi\qquad \mbox{and}\qquad y=y_{q\pm}~.
\ee
Since we are at the pole of the deformed sphere $(\th,\,\phi)$ and at
the pole of the deformed squashed sphere $(y,\,\psi)$ we additionally get
\be
\dot{\phi}=0\qquad \mbox{and}\qquad \dot{\psi}=0~.
\ee
It turns outs that the equations \eq{bpseq1} and the last equation of \eq{bpseq2} are satisfied as well as the first equation of motion \eq{eom1}. The equation \eq{eom2} however introduces a relation between the deformation parameter $\gh$ and the coordinate $y$, and this implies that a BPS string solution for a fixed deformation parameter $\gh$, exist only in a particular Sasaki-Einstein manifold.
The exact equation is
\be
y=\pm\frac{\sqrt{3}}{2 \gh}~,
\ee
which using \eq{equationsbps} implies
\be\label{bg}
a=\frac{3 \left(3 \gh\mp\sqrt{3} \right)}{4 \gh^3}~.
\ee
The rest of the equations of motion are satisfied providing that $\ddot{\a}=0$.

The dependence of $a$ on $\gh$ is a peculiar situation which leads to several comments.
For a particular Sasaki-Einstein manifold, the BPS string solutions exist only for some fixed values of $\gh$. By solving the equation \eq{bg} for $\gh$ we can find the discrete values of $\gh$ that are allowed by giving appropriate values to $a$, and we plot the results in the Figure 2. Although the curve is very dense and almost continuous, it seems that $\gh$ can not take integer values. Furthermore, the constraint \eq{acon} introduces a low bound in the deformation parameter $\gh>\sqrt{3}/3$. The existence of this lower bound explains why these solutions do not reduce to the undeformed ones for $\gh\rightarrow 0$.

In the corresponding field theory, the dual BPS operators to the BPS strings found here, should exist only for specific values of the deformation parameter $\beta$, which should be in agreement with the supergravity calculation. It is natural to expect that in the field theory some special operators exists such that they are non-BPS for generic values of $\beta$ but becomes BPS when \eq{bg} holds.
\begin{figure}[t]
\centerline{\includegraphics[width=80mm]{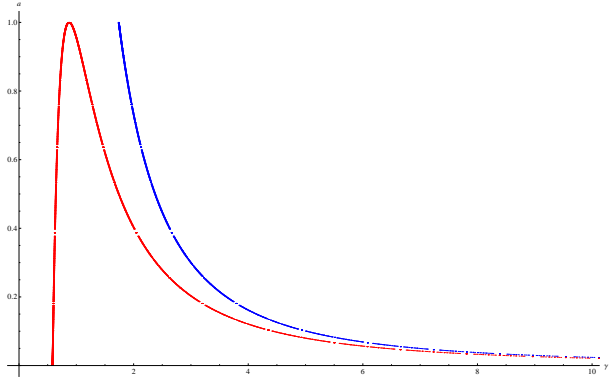}}
\caption{\small{ The discrete values of the deformation parameter $\gh$, for the corresponding $Y^{p,\,q}$ manifold where the BPS string solutions exist, according to equation \eq{bg}. The curve is very dense. Nevertheless, it seems that no integer values for $\gh$ are allowed.}}
\end{figure}

Moreover we observe that the deformed BPS string solutions live where the two $U(1)$ circles shrink to zero size and the corresponding two torus collapses.
The image of the moment map for $C(Y^{p,\,q})$ is a  convex rational polyhedral cone in  $\mathbb{R}^3$, and the edges of the cone can be found by finding the submanifolds which are fixed under a $\mathbb{T}^2$ action. The facets of the polyhedron have $\mathbb{T}^2$ fibration, where on the edges it degenerates more to $\mathbb{T}^1$. Hence the images of the submanifolds where the BPS string lives, correspond to the edges of this polyhedral cone.

In summary the BPS string solutions in the $\beta$-deformed Sasaki-Einstein gauge/ gravity dualities live where the two $U(1)$ circles shrink to zero and in the $\mathbb{T}^3$ fibration description on the edges of the polyhedron. Furthermore, for each $Y^{p,\,q}$ manifold, these string solutions exist only for some specific  values of the deformation parameter which we have found.

\section{Non-BPS string solutions in the $\beta$-deformed $Y^{p,q}$ background }

In this section we study strings that are allowed to move in $Y^{p,q}$ along some of the three $U(1)$ angles $\a,\,\phi,\,\psi$ and are at rest along all the other directions. The most general ansatz, for the type of strings we examine is
\ben\label{ansatz1}
&&\a=\o_1\t+m_1 \s~,\qquad\phi=\o_2\t+m_2 \s~,\qquad\psi=\o_3\t+m_3 \s~,\\\nonumber
&&\theta=\theta_0~, \hspace{2.5cm} y=y_0~,
\een
where $\theta_0$ and $y_0$ are constants and their exact values should be chosen to be consistent with the solutions of the equations of motion and the Virasoro constraints. Moreover, due to the periodicity condition in the global coordinates of the manifold on $\sigma$, the winding numbers have to be integers.
Finally, for the linear dependence on $\t,\,\s$ of the ansatz \eq{ansatz1}, the equations of motion \eq{eom3}, \eq{eom4} and \eq{eom5}  for the $U(1)$ angles are trivially satisfied.

In the following sections, we are going to consider some special cases of the ansatz
\eq{ansatz1} for the string motion.

\subsection{One angle solutions}

In this section we examine the simplest motion for a string, where it is allowed to move only along $U(1)$ direction.
In the undeformed manifold, it has been found using a similar coordinate system, that there exist only one acceptable solution which has single spin.
Here we also try to investigate whether the situation is similar or the deformation parameter $\gh$ enters in the equations of motion in a way that new deformed solutions appear along the directions where string motion is not allowed in the undeformed case.
Moreover we examine how the energy-spin relations are modified.

It is known that only point-like strings are allowed to move along a single $U(1)$ direction. This can be seen by considering an extended string along a $U(1)$ direction and solving the first Virasoro constraint, to see that does not give any acceptable solutions. So we continue here to study the point-like strings.

We start by examining a point-like string moving according to $\a=\o_1 \t$. Unlike the undeformed case we have two non trivially satisfied equations \eq{eom1}, \eq{eom2} which are
\ben
w \partial_\th G  \o_1^2=0\label{p21}\\
(w \partial_y G +G W)\o_1^2=0\label{p22}
\een
The equation \eq{p21} can be written as
\ben\label{pp21}
\frac{72 \o_1^2 \gh^2 \left(y^2-a\right) \left((3-y) y^2+a (1-3 y)\right) s_{2\th}}{\left(-18+a \gh^2 (3 y-5)+y \left(18+\g^2 y (9-7 y)\right)+\g^2 \left((3-y) y^2+a (1-3 y)\right) c_{2\th}\right)^2}=0
\een
and it is proportional to $\gh$ so in the undeformed case it is satisfied trivially.
Supposing $\gh \neq 0$ one gets the following two real solutions from this equation:
\ben\label{ya11}
y=1+\frac{\left(1-a\right)^{1/3}\left(\left(1-a\right)^{1/3}+\left(1+\sqrt{a}\right)^{2/3}\right)}
{\left(1+\sqrt{a}\right)^{1/3}}\quad\mbox{or} \quad \th=0~.
\een
It is obvious that the solution for $y$ is not acceptable since $y> 1$. Hence the only option is to consider $\theta=0$ and go to the pole of the round sphere. Then the equation \eq{p22} becomes
\ben\label{p23}
18\, \frac{4 \gh^2 y^4+3 \left(3-4 a \gh^2\right) y^2+2 \left(4 a \gh^2 -9\right) y+9 a}{\left(9-9 y+2 \gh^2 \left(a+y^2 (2 y-3)\right)\right)^2}=0
\een
and has four lengthy solutions $y(\gh,\,a)$. The two of them are outside the desirable range, but the other two are acceptable. In Figure 3 we plot the acceptable solutions where we call $y_1$ the one valid for small $\gh$, and $y_2$ the one valid for bigger $\gh$.  When we compare these solutions to the undeformed one, we see that at the limit $\gh\rightarrow 0$ the solution $y_1$ approaches the undeformed one. For the same limit the solution $y_2$ is not defined and can not reproduce the undeformed result.
\begin{figure}[t]
\centerline{\includegraphics[width=70mm]{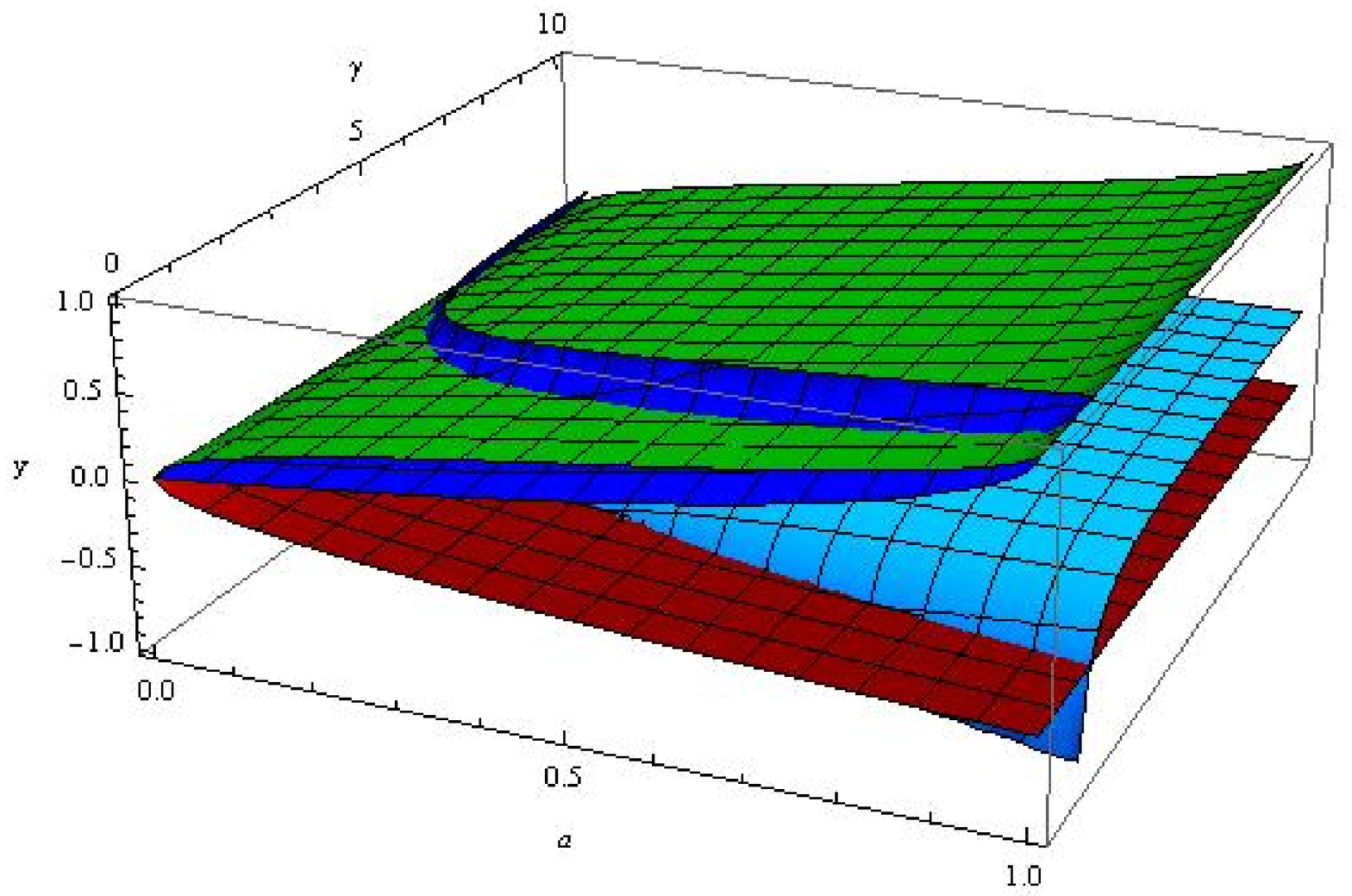}\quad \includegraphics[width=70mm]{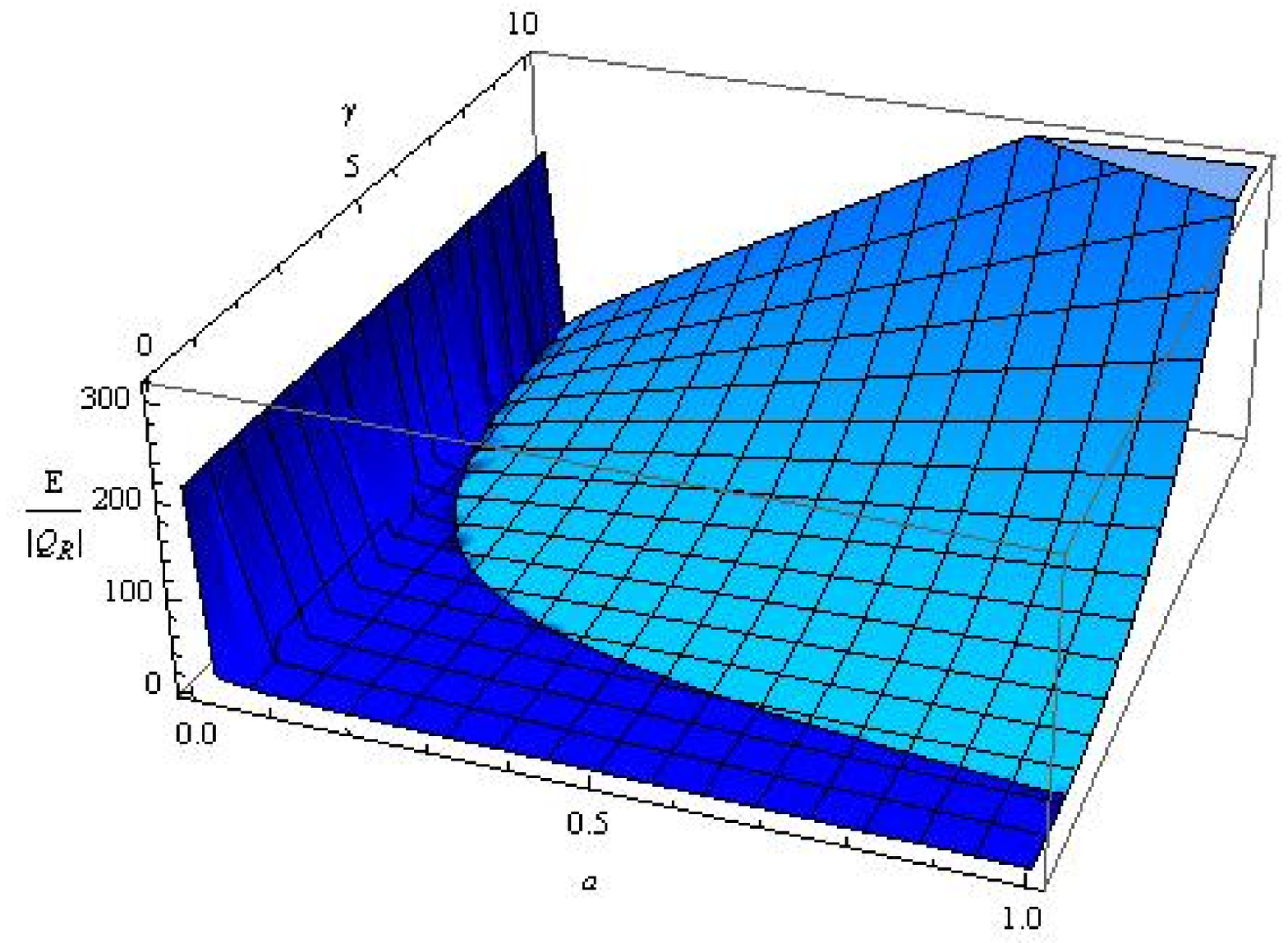}}
\caption{\small{In the first plot are presented the two acceptable solutions $y_1$ and $y_2$ that solve the equations \eq{p21} and \eq{p22} and satisfy for a wide range of values $(a,\,\gh)$ the constraint \eq{yqy}. The color convention is similar to the Figure 1 and we keep it for the rest of the paper. The solutions $y_{q-},\,y_{q+}$ are plotted with red and green respectively. In the second plot, we see the energy divided by $|Q_R|$ as a function of $a$ and $\gh$. The two different solutions, generate different behavior in the energy function.
}}
\end{figure}

The next step is calculate the energy-spin equation of these solutions. The conserved momenta are
\ben
J_a=\sqrt{\l}\, G\, w\, \o_1\,,\qquad J_\psi=-J_\phi=\sqrt{\l} G\, w \,f\,\o_1\,,\qquad J_a=J_{tot}\,,
\een
while the second Virasoro constraint gives for the energy
\ben\label{sole1}
E=\sqrt{\l} \sqrt{G w \o_1^2}
\Rightarrow E=\sqrt{\frac{1}{w}\left(1+\gh^2 Z\right)J_\a^2}~,\quad\mbox{or}\quad E=\frac{3 }{|6f-1|\sqrt{G w}}|Q_R|~,
\een
where $y=y_{1,\,2}(\gh,\,a)$ should be inserted in the relevant functions. The exact behavior of the energy divided by $|Q_R|$ is shown in Figure 3. We observe that the energy of the deformed solutions is modified by an overall factor of $\sqrt{G^{-1}}$ compared to the undeformed one\cite{giataganas3}.
Moreover, one can see that in that case the energy over $|Q_R|$ is greater than the relevant undeformed result.
As expected, at the limit  $\gh\rightarrow 0$ the deformed  and undeformed energies become equal, since the equation \eq{sole1} becomes equal to the undeformed one found in \cite{giataganas3}, and the solution $y_1$ which enters in the energy relation through the functions $w$ and $f$ tends to the relevant undeformed solution at this limit.

In the undeformed manifold it has been found that the point-like string is impossible to move along each one of the other two $U(1)$ directions \cite{giataganas3}, since the solutions for $y$ were not acceptable.
To investigate the situation in the deformed case we consider the linear ansatz $\phi=\o_2 t$, where we get two non trivially satisfied equations of motion for $\theta$ and $y$
\ben\label{sol2eom1}
\partial_\theta G\,\Big(\frac{1-y}{6}s_\th^2+(\frac{q}{9}+ w f^2)c_\th^2\Big)+ G\Big(\frac{1-y}{6}-(\frac{q}{9}+ w f^2)\Big)s_{2\th}=0~,\\\label{sol2eom2}
\partial_y G\,\Big(\frac{1-y}{6}s_\th^2+(\frac{q}{9}+ w f^2)c_\th^2\Big)+ G\Big(-\frac{1}{6}s_\th^2+(\frac{Q}{9}+ A_1 )c_\th^2\Big)=0~.
\een
For $\theta\neq 0$ these equations have two complex solutions. For $\theta=0$,  the equation \eq{sol2eom1} is satisfied without any further requirement and from \eq{sol2eom2} we get four real solutions where only one of them satisfies the Sasaki-Einstein constraints:
\be
\th=0~~\mbox{and}~~ y=1-\frac{\left(-9+4 \left(1-a\right) \gh^2+\sqrt{81+8 \gh^2 \left(1-a\right) \left(2 \left(1-a\right) \gh^2-15\right)}\right)^{\frac{1}{2}}}{2 \sqrt{2} \gh}~.\label{sold2}
\ee
In order for $y$ to be real we need to impose
\ben
\gh>\frac{3 \sqrt{3}}{2}\qquad \mbox{and}\qquad 0<a<\frac{4 \gh^2-27}{4 \gh^2}~,
\een
which are actually a bit weaker conditions than the ones  imposed by the constraint \eq{yqy} (Figure 4). 
An important remark here is that $\gh$ is bounded from below. This is expected since in the undeformed background there is no such a solution.
\begin{figure}[t]
\centerline{\includegraphics[width=70mm]{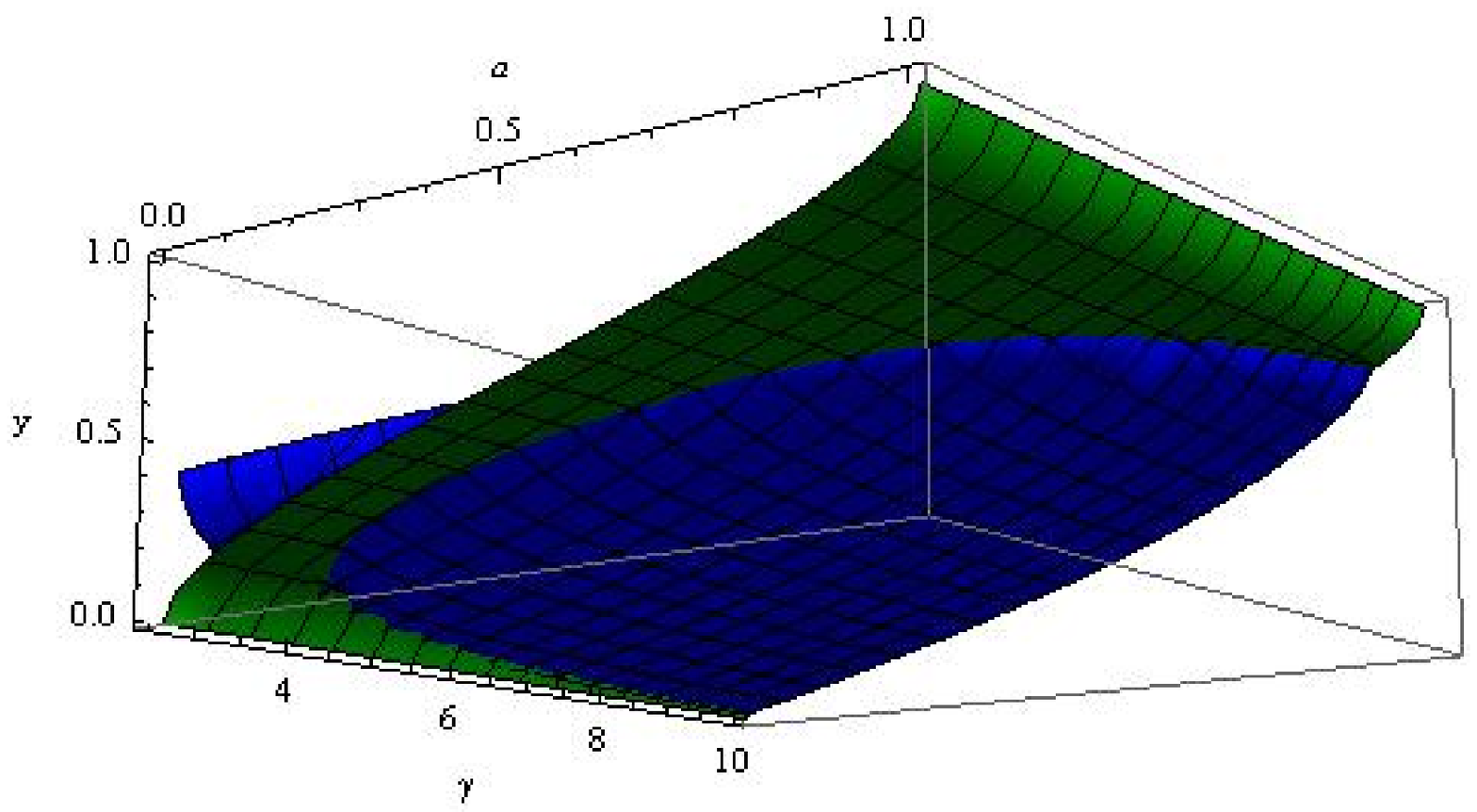}\qquad\includegraphics[width=52mm]{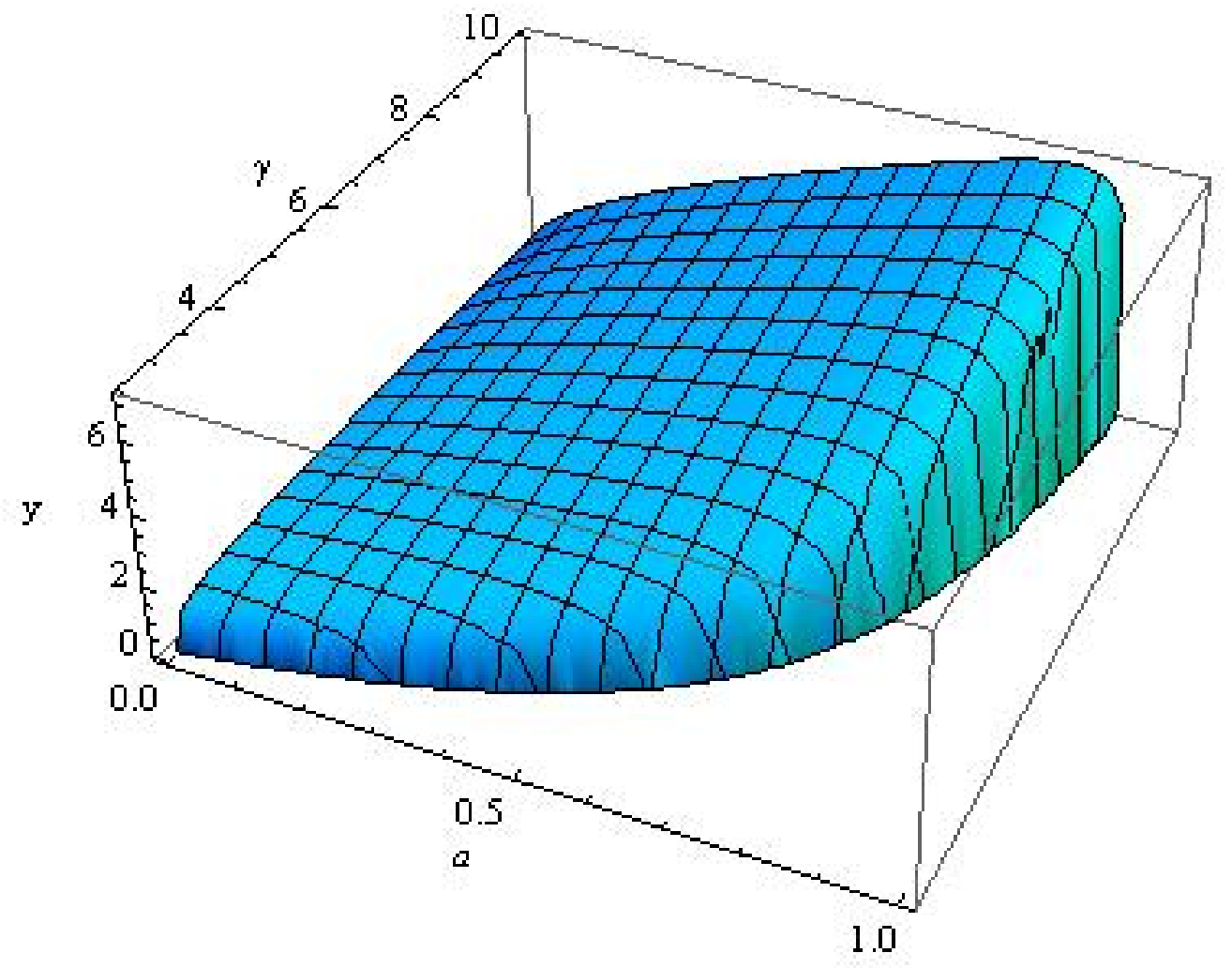}}
\caption{\small{On the left the acceptable solution \eq{sold2} in the range where it is real and the $y_{q+}$. $y$ is between $y_{q+}$ and $y_{q-}$ for a wide range of $(a,\gh)$ values. On the right we draw the energy of a point-like string moving along $\phi$,  divided by $|Q_R|$ as a function of $a$ and $\gh$.}
}
\end{figure}

The conserved momenta in that case are
\ben\label{momentaphi1}
J_\a=-\sqrt{\l} G w f \o_2\,,\qquad  J_\phi=-J_\psi= \sqrt{\l}G(\frac{q}{9}+w f^2)\o_2\,,\qquad J_{tot}=J_\a\,,
\een
and the second Virasoro constraint gives for the energy
\ben
E=\sqrt{\l}\sqrt{G (\frac{q}{9}+w f^2)\o_2^2}~,
\een
which in terms of $Q_R$ becomes
\ben\label{energyphi1}
E=3\left(\frac{q+ 9 w f^2}{G \left(2 q-3 \left(1-6 f\right) f w\right)^2}\right)^\frac{1}{2}|Q_R|~,
\een
and is plotted in Figure 4. Of course the solution for $y$ in \eq{sold2} should be inserted in the above functions. We note that for increasing $\gh$ the energy of the string increases in any $Y^{p,q}$ manifold.

Finally we briefly consider motion along the third $U(1)$ angle with $\psi=\o_3 \t$. The equations of motion become
\ben\label{sol3eom1}
\gh^2 G \partial_\th h+\partial_\th G\left(\frac{q}{9}+ w f^2+ \gh^2 h\right)&=&0~,\\\label{sol3eom2}
G \left(\frac{Q}{9}+ A_1+ \gh^2 \partial_y h\right) +\partial_y G\left(\frac{q}{9}+ w f^2+ \gh^2 h\right)&=&0
\een
and for $\th=0$ are reduced to the equations \eq{sol2eom1} and \eq{sol2eom2} respectively, which have acceptable solution given by \eq{sold2}. The momenta, and the energy spin relation, are given by  \eq{momentaphi1} and \eq{energyphi1} respectively.

Some comments on the non-BPS solutions found in this section are in order. For the point-like string moving along $\phi$ and $\psi$ directions the energy over $|Q_R|$ is increasing as a function of $\gh$. The behavior of the energy of the string for motion along the $\a$ direction is similar for bigger $\gh$. For small $\gh$, where the solution $y_1$ is valid, the energy slightly decreases as $\gh$ increases. This is unusual but happens because the functions that appear in the energy spin relation \eq{sole1} depend on $y$ 
and eventually depend on $\gh$ for this solution. Hence although the overall factor in the dispersion relation is of the form $G^{-1}=1+\gh^2 Z$, where $Z>0$, the energy may decrease or increase with respect to $\gh$.
Moreover, we point out that the solutions for point-like strings moving along $\phi$ or $\psi$ directions
are valid for $\gh$ bounded from below by a finite value, and that is the reason that they do not exist in the undeformed Sasaki-Einstein manifolds.

We continue our analysis in the next section by examining extended string solutions.

\subsection{Extended string solution}

In this section we allow the string to move along the two $U(1)$ directions, we solve the equations of motion and find the dispersion relation which we compare it to the undeformed one found in \cite{giataganas3}. We choose a string that is extended and rotate along the $\a$ direction and it can also move along $\psi$ direction\footnote{The point-like version of this string for appropriate values of the frequencies is the BPS one in the undeformed theory.}. The corresponding ansatz is
\ben
\a=\o_1 \t + m_1\s~,\qquad \psi=\o_3 t \,.
\een
Inserting that to the equations of motion \eq{eom1} and \eq{eom2},  simplify to
\ben\nonumber
&&\partial_\theta G\,\Big(-(\frac{q}{9}+w f^2 +\gh^2 h)\o_3^2+ w(-\o_1^2+m_1^2)  - 2 w f\o_1 \o_3 -2 \gh c_\th \frac{w q}{9} m_1 \o_3\Big)\\\label{p2a1}
&&\hspace{6.8cm}+G\gh \Big(\frac{ 2 w q}{9} s_\th m_1 - \gh \partial_\th h\o_3\Big)\o_3 =0,\\\nonumber
&&\partial_y G\,\Big( -(\frac{q}{9}+w f^2 +\gh^2 h)\o_3^2+ w(-\o_1^2+m_1^2) -2 w f\o_1 \o_3 -2 \gh c_\th \frac{w q}{9} m_1 \o_3\Big) \\\label{p2a2}
&&+G\Big( W (m_1^2- \o_1^2 ) - (\frac{Q}{9}+A_1 +\gh^2 \partial_y h)\o_3^2 -2( A_3 \o_1 + \gh c_\th \frac{A_4}{9}m_1)\o_3\Big)=0,
\een
while the first Virasoro constraint reads
\ben\label{simpl}
G w( \o_1 + f \o_3)m_1=0~.
\een
We start by solving the simplest equation \eq{simpl}, and concentrate to the solution
\ben
y=a~\qquad \mbox{and}\qquad \o_1=\frac{\o_3}{6}~.
\een
To satisfy the condition \eq{yqy}, the parameter $a$ must be constrained inside the following interval
\be\label{a12mikro}
0<a<1/2
\ee
and to make things simpler we choose the angle $\th=\pi/4$.
Then \eq{p2a1} is solved for\footnote{There exist another solution to this equation but by substituting it to the equation \eq{p2a2} gives not acceptable solution for  $a$.}
\be\label{grtele2}
\o_3=\frac{3 \gh\, a (1+a) m_1 }{\sqrt{2} (1-2 a) \left(3+(1-a) a \gh^2\right)}~.
\ee
Then equation \eq{p2a2} simplifies to
\be\label{amtele1}
\frac{3 a m_1^2 \left(6+(a-1) a \left(24+(7+a (7 a-13)) \gh^2\right)\right)}{(1-2 a)^2 (1-a) \left(3+(1-a) a \gh^2\right)^2}=0
\ee
and is a fifth order equation of $a$ which gives one zero solution, two complex solutions and two lengthy real solutions of the form $a(\gh)$. Both of the real solutions satisfy \eq{acon}, but only one of them, which we call $a_1(\gh)$, satisfies the constraint \eq{yqy} (Figure 5).
\begin{figure}[t]
\centerline{\includegraphics[width=70mm]{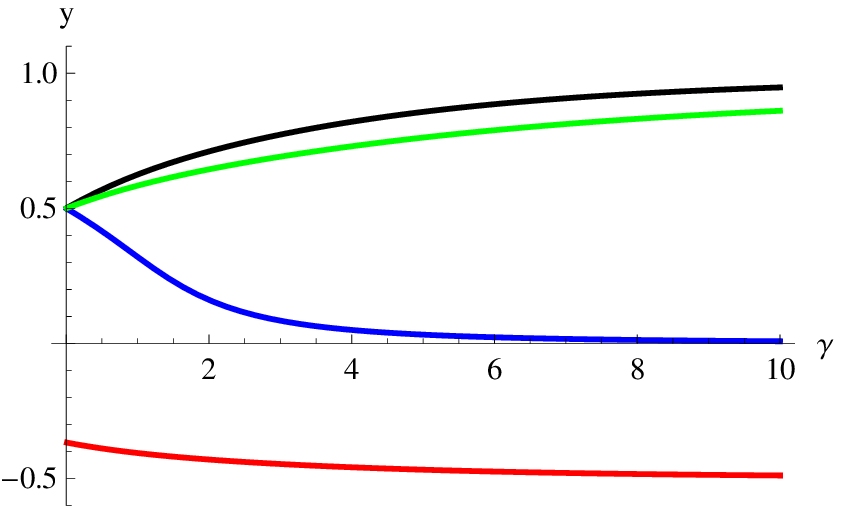}\qquad\includegraphics[width=70mm]{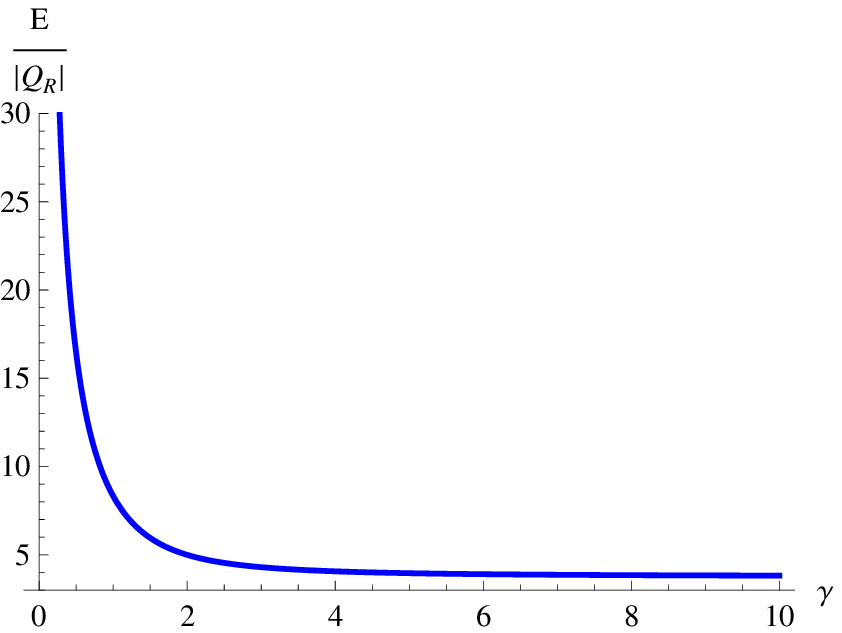}}
\caption{\small{In the first plot are the solutions $y(=a)$ of \eq{p2a2} in  terms of $\gh$ and the functions $y_{q\pm}(\gh)$. We see that only one solution $a_1(\gh)$, plotted with blue color, is between $y_{q\pm}$. The solution $a_2(\gh)$ plotted with black is outside the acceptable range and this is expected since $a>1/2$ and \eq{a12mikro} is not satisfied. The reason that the curves $y_{q\pm}$ do not have the usual form is that  $a_{1,\,2}(\gh)$ have been substituted to $y_{q\pm}(a)$. In the second plot we see the quantity $E/|Q_R|$ vs $\gh$. The energy approaches a constant value quickly. This is expected since the function $a_1(\gh)$ in the same region, has almost already taken a constant value as we see from the first plot.}
}
\end{figure}
The form of these string solutions indicates that for certain values of $a$ the parameter $\gh$ takes an appropriate value given by $a_1(\gh)$. Or equivalently if one deforms a specific Sasaki-Einstein manifold, these string solutions only exist for certain values of $\gh$. Consequently $\gh$ should satisfy the rationality conditions for $a$. Nevertheless, if we did not fix the angle $\theta=\pi/4$ the above solutions would be of the form $a(\th,\gh)$ and they would exist on any manifold and for any deformation parameter $\gh$. Hence in general  the situation is not so constraining as the BPS string solutions found in section 3.

It is clear that the conjugate momenta will depend on one winding number or frequency. We calculate them as
\ben
J_\a=0,\qquad J_\psi=-\frac{\sqrt{2}}{2}J_\phi=\frac{(1-a) a  \gh}{\sqrt{2} \left(3+(1-a) a \gh^2\right)}m_1~,
\een
where  $a=a_1(\gh)$ but  the result  is not written here explicitly in order to keep the relations short. The second Virasoro constraint gives
\be
\k^2=\frac{3 a  \left(12(1-2 a)+(1-a)  (5-7 a) a\gh^2\right)}{2 (1-2 a) \left(3+(1-a) a \gh^2\right)^2} m_1^2~,
\ee
which  in terms of $Q_R$ is written as
\be
E=\sqrt{\frac{15-21 a}{4\left(1-3 a+2 a^2\right)}+\frac{9}{(1-a)^2 a \gh^2}}|Q_R|.
\ee
By inserting $a_1(\gh)$ in the energy relation, the RHS becomes a function of $\gh$ only. This result is very lengthy and for convenience we do not write it here explicitly. Nevertheless, we  plot the result in Figure 5. We see that the energy is reducing fast and approaching asymptotically $\sim 3 .75 Q_R$ already for $\gh\sim 10$. This is expected, since for the same values of $\gh$ the function $a(\gh)$ is approaching asymptotically zero.

We can also  verify our results comparing with the undeformed ones. The deformed solution is valid for $\gh \neq 0$, however we can still study the relevant limit. For $\gh\rightarrow 0 \Rightarrow a\rightarrow 1/2$ and $\o_3=\sqrt{3} m_3$. The solution at this limit is equal to the undeformed solution for $a\rightarrow 1/2$ \cite{giataganas3}.

Let us also consider a specific example of a manifold and compare the deformed and undeformed dispersion relations. By choosing $Y^{2,1}$ which has $a\simeq 0.387327$, we find $\gh=0.652585$ for our string solution.  For these values $E_{deformed}\simeq 12.5614\, Q_R$ and $E_{undeformed}\simeq 12.365 \, Q_R$ which can be found using results in \cite{giataganas3}. The string energy in the undeformed space is smaller than the energy in the deformed one.

To summarize our results, we solve the system of equations by solving firstly the Virasoro constraint \eq{simpl}, and find that its solution relates $y$ to $a$, and $\o_1$ to $\o_3$. We plug this solution to the equation \eq{p2a1} and we express $\o_3$ in terms of $m_1$ linearly \eq{grtele2}. By substituting all these solutions to the equation \eq{p2a2} we get the equation \eq{amtele1} which has an overall dependence on $m_1$. That means that the final solution is of the form $a(\gh)$, where for each value of the deformation parameter the corresponding string solution exist only in a specific $Y^{p,q}$ manifold. If we did not fix the angle $\th$ the solution would have the form $a(\gh,\,\th)$, and then we could have some freedom to choose different manifolds that our string solutions are valid for a single value of deformation parameter.  We have also verified that if we consider the limit $\gh\rightarrow 0$, the solution reduced to the undeformed one found in \cite{giataganas3}.
Finally one can also check that the energy of the string in the deformed space at the limit $\gh\rightarrow 0$ corresponding to $a\rightarrow 0.5$, diverges with the same speed as the energy of the corresponding string in the undeformed background.

\section{Discussions}

In this paper we have found the BPS string solutions in the $\beta$-deformed Sasaki-Einstein backgrounds. These strings live in the submanifold where the two $U(1)$ circles shrink to zero. In the corresponding $\mathbb{T}^3$ fibration over a rational polyhedron description, this is equivalent of saying that the strings live on the edges of the polyhedron where the $\mathbb{T}^3$ fibration degenerates to $\mathbb{T}^1$.
This result is in analogy with the dual giant gravitons examined in the same backgrounds \cite{zaffaroni07}. For example, there it has been found that the BPS D3 dual giant graviton live on the edges of the polyhedron and as expected rotates along the Reeb vector.

Moreover, for the BPS solutions we found, it appears that there is a relation \eq{bg} between the deformation parameter $\gh$ and the Sasaki-Einstein parameter $a$. This means that in a specific Sasaki-Einstein manifold, say $Y^{p_1,q_1}$, the BPS string solutions exist only for
specific values of the deformation parameter $\gh$. In other words, for a fixed value of $\gh$, the BPS string solutions exist only in some particular $Y^{p_1,q_1}$ manifold.

We also examined non-BPS string solutions and derived their dispersion relations. We found that their energy divided by the R-charge in most of the cases is an increasing function of $\gh$. However, at least for one case of the point-like string moving along the fibre direction $\a$ the energy is slightly decreasing with respect to $\gh$. Furthermore, for all of the solutions we found, the energy of the strings over the R-charge, is greater for strings moving in the marginally  deformed background compared to the corresponding strings moving in the undeformed one.

Moreover, we saw that there exist solutions, that are valid only for values of  $\gh$  bounded from below. The corresponding undeformed solutions which are obtained for $\gh\rightarrow 0$  are complex and hence not accepted. In these cases the deformation parameter enter in the equations in a way that allows the deformed solutions to become real for some of its values.

By considering the limit $\gh\rightarrow 0$ on the equations of motion of any $\beta$-deformed background we obtain the undeformed equations and the undeformed solutions. However, by considering the limit $\gh\rightarrow 0$ directly to the solutions of the equations of motion  we do not always get the full undeformed solution, since there can be deformed equations that are proportional to $\gh$.
This equation is of the form $\gh^2 g(\gh,a,\o,m)=0$ and solving it for $\gh \neq 0$ could give a further constraint to the system of equations that is independent of $\gamma$.  If we solve the system of equations that contains the above equation for $\gh \neq 0$, we obtain a solution that for $\gh \rightarrow 0$ gives only a part of the full undeformed solution. Hence one should be careful at which stage takes the limit of $\gh\rightarrow 0$. An example, is the case of point-like string moving along the $a$ direction, where part of the deformed solution is $\theta=0$ which comes from the equation \eq{pp21} that satisfied trivially in the undeformed background for $\gh=0$. However, by considering the limit $\gh=0$ in the initial equations of motion we reproduce the string solution of the undeformed background, which is valid for any $\theta$. In Appendix C, we have a further discussion on this remark.

As a continuation of this work it would be very interesting to reproduce  our results in the dual field theory. Especially to find the dual operators that correspond to the BPS solutions we found. It is natural to think that in the field theory, our results imply that some special operators exists such that are non-BPS for generic values of $\beta$ but becomes BPS and correspond to our BPS string solutions when \eq{bg} holds.  One can also consider the limit that the R-charge is very large, which corresponds to a string with a large angular momentum. Both the supergravity and the field theory analysis in this limit should be doable.

Here we use a single parameter $\beta$-deformation in a way that preserves the $\cN=1$ supersymmetry. One could also make a multi-$\beta$-deformation. These multi-deformed backgrounds are obtained by performing three TsT transformations to the three different pairs of the $U(1)$ angles. Hence the resulting backgrounds depend on three parameters $\gh_i$, each one corresponding to a single TsT transformation. They do not preserve any supersymmetry, so any studies in the context of AdS/CFT are very interesting.
It would be interesting to consider again BPS-like minimizing conditions, and find whether or not the strings still leave on the edges of the polyhedron. It is possible that their energy equation in the multi $\beta$-deformed background remains similar to \eq{kfinal} but gives three copies of the last term of this equation, one for each $\gh_i$. If this is the case, the 'BPS' strings will still live where the two circles shrink to zero, and the only modification will appear in the equation \eq{bg} which should possibly relate the manifold parameter $a$ with all the deformation parameters $\gh_i$.
At least, for the giant gravitons in multi $\beta$-deformed toric backgrounds if one does the calculations, finds that the 'BPS' conditions force the dual giant gravitons to live where the two $U(1)$ circles shrink to zero as in the case of the single $\beta$ deformation.

\textbf{Acknowledgements:} I would like to thank Robert de Mello Koch and Peter Ronne for useful discussions and Sangmin Lee for useful discussions and correspondence.
I would also like to thank Michael Haack and Dieter Lust for hospitality in Arlond Sommerfeld Center for theoretical physics where during my visit, part of this work was done. The research of the author is supported by a SARChI postdoctoral fellowship.

\startappendix
\Appendix{Equations of motion and conserved quantities}

In this section we derive the equations of motion, the Virasoro constraints and the conjugate momenta in the $\beta$-deformed $Y^{p,q}$ background.
As in the undeformed case of \cite{giataganas3, giataganas4} the non-BPS strings we consider are allowed to move on a circle of the deformed round sphere $S^2$ parameterized by the coordinate $\phi$, while $\theta$ kept fixed. On the deformed squashed sphere the string can move on its azimuthal coordinate $\psi$, and sit at a constant value $y$ between the north and south poles. This value will be chosen by solving the equations of motion. Finally, the string can move on the principle $S^1$ bundle over $B$ parameterized by $\a$. Notice that each of the directions that the string is allowed to spin has a $U(1)$ symmetry.
The global time is expressed through the world-sheet time as $t=\k \t$, and the string is localized at the point
in the bulk of $AdS$. The Polyakov action in the conformal gauge is given by
\ben\nonumber
S&=&-\frac{\sqrt\l}{4 \pi}\int d\t d\s \,\bigg[ \dot{t}^2-t'{}^2-\frac{1-y}{6}\left(\dot{\theta}^2-\theta'{}^2\right)-\frac{1}{w q}\left(\dot{y}^2-y'{}^2\right) \\\nonumber
&&\qquad\quad+G\Big[w\left(-\dot{\a}^2+\a'{}^2\right)-\left(\frac{1-y}{6}s_\th^2+\left(\frac{q}{9}+w f^2 \right)c_\th^2\right)\left(\dot{\phi}^2-\phi'{}^2\right)\\\nonumber
&& \qquad\quad -\left(\frac{q}{9}+w f^2+\hat{\g}^2 h\right)\left(\dot{\psi}^2-\psi'{}^2\right)
+2 c_\th\left(\frac{q}{9}+w f^2\right)\left(\dot{\psi}\dot{\phi}-\psi'{}\phi'{}\right)\\\nonumber
&&\qquad\quad-
2 w f\left(\dot{\a}\dot{\psi}- \a'{}\psi'{}\right)+2 w f c_\th\left(\dot{\a}\dot{\phi}-\a'{}\phi'{}\right)\Big]+2G \hat{\g} \Big[-Z\left(\dot{\a}\phi'{}-\a'{}\dot{\phi}\right)\\\label{actiond}
&&\qquad\quad+\frac{w q}{9}c_\th\left(\dot{\a}\psi'{}-\a'{}\dot{\psi}\right)
+\frac{f (a-y^2)}{3}s_\th^2\left(\dot{\phi}\psi'{}-\phi'{}\dot{\psi}\right)\Big]~.
\een
The classical equations of motion for $\th$ and $y$ are: 
\ben\nonumber
&&G\big[\frac{1-y}{6}\left(-\dot{\phi}^2+\phi'{}^2\right)s_{2\theta} + \left( \frac{q}{9}+ w f^2\right)\left(s_{2\theta}\left(\dot{\phi}^2-\phi'{}^2\right)-2 s_\theta\left(\dot{\psi}\dot{\phi}-\psi'{}\phi'{}\right)\right)\\\nonumber
&&-\gh^2 \partial_\th h\left(\dot{\psi}^2-\psi'{}^2\right)-2 w f s_\theta\left(\dot{\a}\dot{\phi}-\a'{}\phi'{}\right)\big]+\partial_\th G L_1\\\nonumber
&&+2 G \gh \left[-\partial_\th Z\left(\dot{\a}\phi'{}-\a'{}\dot{\phi}\right)+\frac{\left(a-y^2\right) f s_{2\th}}{3}\left(\dot{\phi}\psi'{}-\phi'{}\dot{\psi}\right)-
\frac{w q s_\th}{9}\left(\dot{\a}\psi'{}-\a'{}\dot{\psi}\right)\right]\\
&&\qquad\qquad\qquad\qquad\qquad\qquad\qquad\qquad
+\partial_\t\left(\frac{1-y}{3}\dot{\th}\right)-\partial_\s\left(\frac{1-y}{3}\th'\right)=0~, \label{eom1}
\een
\ben\nonumber
&&\frac{1}{6}\left(\dot{\theta}^2-\theta'{}^2\right)-A_2\left(\dot{y}^2-y'{}^2\right)+
G\Big[\frac{s_\theta^2}{6}\left(\dot{\phi}^2-\phi'{}^2\right)+\left(\frac{Q}{9}+ A_1 \right)\Big(c_\theta^2\left(-\dot{\phi}^2+\phi'{}^2\right)\\\nonumber
&&-\dot{\psi}^2+\psi'{}^2+2 c_\theta\left(\dot{\psi}\dot{\phi}-\psi'{}\phi'{}\right)\Big)
-\gh^2 \partial_y h\left(\dot{\psi}^2-\psi'{}^2\right)-W\left(\dot{\a}^2-\a'{}^2\right)\\\nonumber
&&-2 A_3\left(\dot{\a}\dot{\psi}- \a'{}\psi'{}-c_\theta\left(\dot{\a}\dot{\phi}-a'{}\phi'{}\right)\right)\Big]
+\partial_y G L_1\\\nonumber
&&+2 G\gh\Big[-\partial_y Z\left(\dot{\a}\phi'{}-\a'{}\dot{\phi}\right)
+\frac{A_4}{9}c_\th\left(\dot{\a}\psi'{}-\a'{}\dot{\psi}\right)+
\frac{\partial_y\left(f\left(a-y^2\right)\right)}{3}s_\th^2\left(\dot{\phi}\psi'{}
-\phi'{}\dot{\psi}\right)\Big]\\
&&\qquad\qquad\qquad\qquad\qquad\qquad\qquad\qquad\qquad\quad+\partial_\t\left(\frac{2}{w q}\dot{y}\right)-\partial_\s\left(\frac{2}{w q}y'\right)=0~,\label{eom2}
\een
while for the three $U(1)$ angles $a,\,\phi$ and $\psi$ are:
\ben
\partial_{\b}\big[ \g^{\b\d}G\left(w \partial_\d \a +w f \left( \partial_\d \psi-c_\theta\partial_\d \phi\right)\right)- G \gh \e^{\b\d}\big(Z \partial_\d \phi- \frac{w q}{9}c_\th \partial_\d\psi \big)\big] =0~,\label{eom3}
\een
\ben\nonumber
&&\partial_\b\big[\g^{\b\d}G \left(\frac{1-y}{6}s_\theta^2\partial_\d \phi+\left(\frac{q}{9}+w f^2\right)\left(c_\theta^2 \partial_\d \phi-c_\theta\partial_\d\psi\right)
- w f c_\theta \partial_\d a\right)\\
&&\qquad\qquad\qquad\qquad\qquad\quad\quad+ G \gh \e^{\b\d}\left(Z \partial_\d\a+\frac{f (a-y^2)}{3}s_\th^2 \partial_\d\psi\right)  \big]=0~,\label{eom4}\\\nonumber
&&\partial_\b\big[\g^{\b\d}G\left(\left(\frac{q}{9}+w f^2\right)\left(\partial_\d \psi-c_\theta\partial_\d \phi\right)+ w f \partial_\d \a+\gh^2 h \partial_\d \psi\right)\\&&
\qquad\qquad\qquad\qquad\qquad\quad- G \gh \e^{\b\d}\left(\frac{w q}{9}c_\th \partial_\d\a+\frac{f (a-y^2)}{3}s_\th^2 \partial_\d\phi \right)\big]=0~,\label{eom5}
\een
where we have used the conventions
\ben
&&A_1:=\partial_y(w f^2)~,\quad A_2:=\partial_y\left(\left(w q\right)^{-1}\right)~,\quad A_3:=\partial_y(w f)~,\\
&&A_4:=\partial_y(w q)~,\hspace{0.74cm} Q:=\partial_y q~,\hspace{2.05cm} W:=\partial_y{w}
\een
and $L_1$ is just the part of the Lagrangian \eq{actiond} which is multiplied by  $G$.\footnote{$L_1$ is equal to the 2nd, 3rd and the 4rth line of the equation \eq{actiond} divided by $G$.}
The last three equations are written in a more compact form since they are satisfied trivially for our linear ansatz and we have written them for constant $\theta$ and $y$. This statement is general, since linear ansatze on $\t$ and $\s$ for motion along the $U(1)$ directions of the action, will satisfy trivially the corresponding equations of motions for these $U(1)$ angles.

Moreover, we should take into account the Virasoro constraints which are equivalent to setting the components of the energy-momentum tensor  to zero. These read
\ben\nonumber
&&G\big[\frac{1-y}{6}s_\theta^2\dot{\phi}\phi'{}+\left(\frac{q}{9}+ w f^2 \right)\left(c_\theta^2\dot{\phi}\phi'{}+\dot{\psi}\psi'{}
-c_\theta\left(\dot{\phi}\psi'{}+\dot{\psi}\phi'{}\right)\right)
+w \dot{\a}\a'{}\\
&&+\gh^2 h \dot{\psi}\psi'{} +w f \left(\dot{\a}\psi'{}+\dot{\psi}\a'{}-\left(\dot{\a}\phi'{}+
\dot{\phi}\a'{}\right)c_\theta\right)\big]+\frac{1-y}{6}\dot{\th}\th'{}+\frac{1}{w q} \dot{y}y'{}=0,\label{vc1}\\\nonumber
&&\k^2=G\big[\frac{1-y}{6}s_\theta^2\left(\dot{\phi}^2+ \phi'{}^2\right)+ w \left(\dot{\a}^2+ \a'{}^2\right)+2 w f \left(\dot{\a}\dot{\psi}+\a'{}\psi'{}-\left(\dot{\a}\dot{\phi}+\a'{}\phi'{}\right)c_\theta\right)
\\\nonumber
&&+
\left(\frac{q}{9}+ w f^2 \right)\left(c_\theta^2\left(\dot{\phi}^2+ \phi'{}^2\right)+\dot{\psi}^2+ \psi'{}^2
-2 c_\theta\left(\dot{\phi}\dot{\psi}+\phi'{}\psi'{}\right)\right)+\gh^2 h \left(\dot{\psi}^2+\psi'{}^2\right)\big]
\\
&&\qquad\qquad\qquad\qquad\qquad\qquad\qquad\qquad
+\frac{1-y}{6}\left(\dot{\theta}^2+\theta'{}^2\right)+\frac{1}{w q} \left(\dot{y}^2+y'{}^2\right) .
\label{vc2}
\een
In the above equations we also include the terms that are corresponding to a non constant $\theta$ and $y$ because we need them to examine the deformed BPS condition in section 3.

As in the undeformed case the symmetry of $Y^{p,q}$ admits three conserved charges which are the angular momenta corresponding to strings rotating along the $\a,\,\phi$ and $\psi$ directions. Furthermore, the classical energy is also conserved, encoding the translational invariance along $t$. We present below these conserved quantities in a general form :
\ben\label{1e}
E&=&\frac{\sqrt{\l}}{2 \pi}\int_{0}^{2 \pi}d \sigma\, \k ~,
\\\nonumber
J_\a&=&\frac{\sqrt{\l}}{2 \pi}\int_{0}^{2 \pi}d \sigma\,G\big[\left(w \dot{a}-  w f c_\theta \dot{\phi} +w f \dot{\psi}\right)
+\gh\left(Z \phi'{}-\frac{w q}{9} c_\th \psi'{}\right)\big]~, \label{1ja}
\\\nonumber
J_\phi&=&\frac{\sqrt{\l}}{2 \pi}\int_{0}^{2 \pi}d \sigma\,G\Big[ -w f c_\theta \dot{a}+\left(\frac{1-y}{6}s_\theta^2  + \frac{q}{9}c_\theta^2 +w f^2 c_\theta^2\right)\dot{\phi}\\ \label{1jphi}
&&\qquad\qquad\quad\quad-\left(\frac{q}{9} +w f^2\right)c_\theta \dot{\psi}-
\gh\left(Z \a'{}+\frac{f (a-y^2)}{3}s_\th^2 \psi'{}\right)
\Big]~,\\\nonumber
J_{\psi}&=&\frac{\sqrt{\l}}{2 \pi}\int_{0}^{2 \pi}d \sigma\,G\Big[ w f \dot{a}-\left(\frac{q}{9} +w f^2\right)c_\th \dot{\phi}+\left(\frac{q}{9} +w f^2+\gh^2 h\right)\dot{\psi}\\ \label{1jpsi}
&&\qquad\qquad\qquad\qquad\qquad\qquad\quad+\gh\left(\frac{w q}{9}c_\th \a'{}+\frac{f (a-y^2)}{3}s_\th^2\phi'{}\right)
\Big]~.
\een
We also define the new quantities $J_{tot}=J_\a+J_\phi+J_\psi$, $\cE=  E /\sqrt{\l}$ and  $\cJ_i = J_i /\sqrt{\l}$. 

\Appendix{Hamiltonian approach for BPS string solutions in marginally deformed toric geometries}

In this Appendix we briefly discuss a different approach to look at some of the properties of the BPS string solutions in the marginally deformed toric geometries.

It is possible to bring the deformed metric of the marginally  deformed toric geometries in the form \cite{zaffaroni07}
\be
ds^2=g_{ij}dy^i dy^j+H(d\psi+\s_\a d\psi^\a)^2+h_{\a\b}d\psi^\a d\psi^\b~,
\ee
with $\psi^a$, $\a,\,\b=1,\,2,$ being the $U(1)$ angles and with the Reeb vector field $\partial/\partial\psi$. Moreover, $H$ and $h_{\a\b}$ depend on $\gh$, are functions of $y^i$ and can be found explicitly. The function $H$ in the notation of the footnote 3, is equal to $G\left(m  g_{33}+n \g^2 h\right)$, where $m$ is a number determined by some normalization an we can set it equal to one. $n$ is a number determined by the Calabi-Yau condition, and is equal to $9$. The corresponding action reads
\be
S=-\frac{\sqrt\l}{2}\int d\t\left(-\dot{t}^2+g_{ij}\dot{y}^i\dot{y}^j+H \left(\dot{\psi}+\s_\a\dot{\psi}^\a\right)^2+h_{\a\b}\dot{\psi}^\a \dot{\psi}^\b\right)~,
\ee
where the conjugate momenta are equal to
\ben
J_\psi&=&\sqrt{\l} H\left(\dot{\psi}+\s_\a\dot{\psi}^a\right)~,\\ J_{\psi^\a}&=&\sqrt{\l} \left[H\s_\a\left( \s_\b\dot{\psi}^\b + \dot{\psi}\right)+ h_{\a\b}\dot{\psi}^\b\right]~.
\een
After some algebra the Hamiltonian can be written in terms of the above momenta as
\be
\cH=\frac{1}{H}J_\psi^2+ h^{\a\b}\left(J_{\psi^\a}-\s_a J_\psi\right)\left(J_{\psi^\b}-\s_b J_\psi\right)+g_{ij}J_{y^i}J_{y^j}~.
\ee
Now one needs to minimize the above expression in order to get the BPS condition, and to satisfy the equations of motion. It turns out that one has at least to consider
\ben
J_{\psi^\a}=\s_\a J_\psi~,\qquad H=1~,\qquad J_{y^i}=0~,
\een
which together with the equations of motion set $Z=0$ and hence restrict the BPS strings to live in the submanifold where the two  $U(1)$ circles shrink. By solving all the equations of motion for the Sasaki-Einstein manifolds explicitly and inserting the exact functions in the metric, one should also find the relation \eq{bg} between the parameter $a$ and the deformation parameter $\gh$ .

\Appendix{Discussion on the $\gh\rightarrow 0$ limit of the deformed solutions}

Usually it is straightforward to see that the deformed solutions in the limit $\gh\rightarrow 0$ reproduce the exact undeformed result. However there are cases, where by considering this limit directly on the deformed solutions and not in the initial equations, only a subspace of the full undeformed solutions is reproduced.

To elaborate the argument further, let us give a simple example by supposing the case where a point-like string moving with a linear dependence on time along a $U(1)$ direction of a 5-dimensional $\beta$-deformed manifold with three $U(1)$ isometries. Then the two non trivial equations of motion for the non $U(1)$ directions parameterized by $y$ and $\th$  take the following form
\ben
\partial_{y,\,\theta} G \,f(y,a,\th,\gh)+ G \partial_{y,\,\theta} f(y,a,\th,\gh)=0\,,
\een
which can be written as
\ben\label{gg1}
f_1(\gh) ~g_{1,\,2}(y,a,\th,\gh)+  \partial_{y,\,\theta} w_{1,\,2}(y,a,\th,\gh)=0~.
\een
Where $f_1(\gh)$ is a function of $\gh$,\footnote{Usually equal to $\gh^2$.} with the property that $f_1(0)=0$ and $g_{1,\,2}$ and $w_{1,\,2}$ two different functions of $(y,a,\th,\gh)$, with  $w_{1,\,2}(y,a,\th,0)\neq 0$~.
For $\gh\neq 0$, the equations \eq{gg1}  give in general a 2-dim solution of the form $y(a,\gh)$ which reduces to undeformed solution for $\gh\rightarrow 0$.

It is possible that the functions $w_{1,\,2}$ is independent of $\th$ or $\a$. By supposing $\th$ independence, \eq{gg1} becomes
\ben\label{gg2a}
f_1(\gh) g_1(y,a,\th,\gh)+  \partial_{y} w_1(y,a,\gh)=0~,\\
f_1(\gh) g_2(y,a,\th,\gh)=0~.\label{gg2b}
\een
For $\gh=0$, the solution of the above system in general is a function of the form  $y_1(a,\th)$  satisfying \eq{gg2a}. For $\gh \neq 0$ the equation \eq{gg2b} gives a new constraint $g_2(y,a,\th,\gh)=0$ and if the solution of this equations does not depend on $\gh$,\footnote{For example the $\gh$ dependence in the function $g_2$ appear only in the denominator.}  will have the form $y_2(a,\th)$. The equation \eq{gg2a} in general gives a solution of the form $y_3(a,\th,\g)$.
Now consider the limit $\gh\rightarrow 0$ for the solutions of the above equations. The solution $y_3$ reduces to the solution $y_1$ which is the full undeformed result. However, since in general there is no reason that $y_1$ should be equal to $y_2$ at this limit, the total solution of the system is a 1-dim curve of the form $y(a)$ which is only a subset of the full undeformed solution.

However if we consider the same limit for $\gh$ in the initial system of equations, the equation \eq{gg2b} is satisfied trivially, and hence the solution of the system is a 2-dim surface where the curve $y(a)$ belongs.
Therefore in this example taking the limit $\gh\rightarrow 0$ on the deformed solution reproduce only a subset of the whole undeformed solution. On the other hand by taking the limit in the initial system of the deformed equations, the equations and their solutions are obviously exactly the same as in the undeformed case.

\Appendix{Some formulas for the Sasaki-Einstein manifolds}

In this Appendix we give some basic formulas we need in the main sections.

The undeformed Sasaki-Einstein metrics $Y^{p,q}$ on $S^2\times S^3$ have the following local form \cite{gauntlett04b}:
\ben
  d s^2 &=& \frac{1-cy}{6}(d\theta^2+\sin^2\theta
      d\phi^2)+\frac{1}{w(y)q(y)}
      d y^2+\frac{q(y)}{9}(d \psi-\cos\theta  d \phi)^2 \nonumber\\  \label{sasakiundef}
      & + &  {w(y)}\left[d \alpha +f(y) (d\psi-\cos\theta
      d \phi)\right]^2.
\een
Generally, it is necessary to have $p/q$ rational in order to have  a complete manifold and the difference between the two roots of $q(y)$ is
\ben\label{yqyq}
y_{q+}-y_{q-}=\frac{3 q }{2 p}~.
\een
If the roots $y_{q+},\,y_{q-}$ are rational we have quasi-regular Sasaki-Einstein manifolds.
However the rationality of $p/q$ can be achieved even in cases that the two roots are irrational which gives irregular Sasaki-Einstein metrics.

The parameter $a$ in terms of $p,\,q$ is
\be\label{apq}
a=\frac{1}{2}-\frac{p^2-3q^2}{4 p^3}\sqrt{4 p^2 -3 q^2}~,
\ee
then the period of $\a$ is given by $2 \pi l$ where
\ben\label{aper}
l=\frac{q}{3q^2-2 p^2+p(4p^2-3 q^2)^{1/2}}~.
\een
The three roots of cubic satisfy
\ben\label{yprop}
y_{q+}+y_{q-}+y_3=3/2,\quad
y_{q+}y_{q-}+y_{q+}y_3+y_{q-}y_3=0,\quad
2 y_{q+}y_{q-}y_3=-a
\een
and also can be expressed in terms of $p,\,q$
\ben
y_{q\pm}=\frac{1}{4p}(2 p \pm 3 q-\sqrt{4 p^2-3 q^2}),\quad
y_3=\frac{1}{4p}(2 p +2\sqrt{4 p^2-3 q^2})
\een
and the period of $a$ \eq{aper}, can be rewritten in a more compact form
\be
l=-\frac{q}{4 p^2 y_{q+} y_{q-}} ,
\ee
which is always positive since $y_{q-}$ is negative. The volume $Y^{p,q}$ is given by
\ben
Vol(Y^{p,q})=\frac{q(2 p +\sqrt{4 p^2-3 q^2})l \pi^3}{3 p^2}
\een
and is bounded by
\be
Vol(T^{1,1}/\mbox{\cZ}_p)>Vol(Y^{p,q})>Vol(S^5/\mbox{\cZ}_2 \times \mbox{\cZ}_p).
\ee

\end{document}